\documentclass[seceq]{ptptex}

\usepackage{graphicx}
\newcommand{\Sfl}[1]{\ooalign{\hfil/\hfil\crcr $#1$}}

\notypesetlogo                       %comment in if to eliminate PTPTeX 
\title{
Disappearance of quasi-fermions in the strongly coupled plasma\\
from the Schwinger-Dyson equation\\ with in-medium gauge boson propagator%
}

\author{%       %Use \scshape  for the family name
Masayasu \textsc{Harada}$^{}$\footnote{E-mail: harada@hken.phys.nagoya-u.ac.jp} 
and Shunji \textsc{Yoshimoto}$^{}$\footnote{E-mail: yoshimoto@hken.phys.nagoya-u.ac.jp} 
}

\inst{%     %Affiliation, neglected when [addenda] or [errata]
Department of Physics, Nagoya University, Nagoya 464-8602, Japan
}

\abst{%         %this abstract is neglected when [addenda] or [errata]
We non-perturbatively study the fermion spectrum 
in the chiral symmetric phase focusing on the %coupling dependence and the
effects of in-medium corrections for gauge boson.
The fermion spectrum is derived by solving the Schwinger-Dyson equation (SDE)
with ladder approximation on the real time axis. 
It is shown that the peak of the fermion spectral function is broadened by 
in-medium effects for gauge boson compared with the peak obtained with the
tree gauge boson propagator. 
The peak becomes much broader as the value of the gauge coupling increases.
This broadening is caused by multiple scatterings of fermions and gauge bosons
included through the non-perturbative resummation done by the SDE.
In particular, the Landau damping of gauge boson propagator plays an important
role in the broadening.
Our results show no clear peak in the strong coupling region, 
implying the disappearance of quasi-fermions in the strongly coupled plasma.
This indicates
that quasi-particle picture may be no longer valid in the strongly coupled
QGP. 
}

\begin{document}

\maketitle

 %%%%%%%%%%%%%%%%%%%%%%%%%%%%%%%%%%%%%%%%%%%%%%%%%%%%%%%%%%%%%%%%%%%%
\section{Introduction}
\label{sec:introduction}
In the quark-gluon plasma (QGP) 
quarks and gluons, existing as elementary degrees of freedom, have
different spectra from those for the non-interacting quarks and gluons. 
In the weak coupling limit ($g\ll 1$), employing the Hard Thermal Loop (HTL)
approximation, it is
shown\cite{Weldon:1982bn,LeBellac}
that light quarks (such as $u$ and $d$) and gluons
have poles with masses proportional to the temperature 
$T$ and the gauge coupling $g$
but
no widths.

Recent experimental studies carried out at RHIC\cite{Arsene:2004fa}, on the
other hand,  
suggests that the QGP near the phase
transition is a strongly interacting system.
This result seems consistent with the findings of recent studies
employing lattice QCD,
which show that the lowest charmonium state survives at $T$
higher than the critical temperature ($T_c$)\cite{Asakawa:2003re}.

For the thermal mass of quark near $T_c$, 
several analyses show $M_{\rm quark}/T \sim 1$ based on 
one-loop calculation taking the gluon condensate into
account\cite{Schaefer:1998wd}, 
Brueckner-type many-body scheme and data from the heavy quark potential
in the lattice QCD\cite{Mannarelli:2005pz},
Nambu--Jona-Lasinio model\cite{Kitazawa:2005mp}.
%While a quenched lattice QCD using the maximum entropy
%method\cite{Petreczky:2001yp} shows that the thermal mass of quark is 
%$M/T=3.9$ at $T=1.5T_c$.
In the work done in Refs.~\citen{Harada:2007gg} and \citen{Harada:2008vk},
an analysis employing the Schwinger-Dyson equation (SDE) is carried out to
show that the thermal mass of fermion in the very strong coupling
region saturates at $M_{\rm quark}/T\sim 1$, which is consistent with the result by a
quenched lattice QCD with two-pole fitting\cite{Karsch:2007wc}.

It is important to study not only the thermal mass but also the entire
spectral function for clarifying the properties of quarks in medium. 
In Ref.~\citen{Mannarelli:2005pz}, the imaginary part of the quark self-energy
is shown to be on the order of the thermal mass. 
Analyses in Refs.~\citen{Harada:2007gg} and \citen{Harada:2008vk} show that
peaks of the spectral function become broad in the strong coupling region.
The SDE used in Ref.~\citen{Harada:2007gg} is solved with 
the tree gauge boson propagator, 
which shows the broadening of the peaks 
with increasing coupling. 
This is understood from the
fact that the probability of gauge boson emission and absorption from a
fermion increases rapidly with the coupling.

In the present work, we investigate the effects 
of medium modified gauge boson on the fermion spectral function. 
We employ the HTL corrected gauge boson propagator to include the 
characteristic properties of collective excitations of the gauge boson, 
i.e. quasi-particle poles and the Landau damping. 
We will pay special attention to the effect of Landau damping and show that it
gives further broadening owing to multiple scatterings with fermions
and gauge bosons,
resulting no clear peak in the spectral function in the strong coupling
region.

The contents of this paper are as follows.
In section \ref{sec:SDE}, 
we introduce the SDE on the real time axis as an iterative equation for
the fermion spectral function\cite{Harada:2008vk}.
In section \ref{sec:NR}, the fermion spectral function obtained in this way
are shown focusing on the effects of medium corrected gauge
boson. 
Section \ref{sec:summary} is devoted to a brief summary and 
discussions. 
In appendices, we show the phase structure in our frame work, 
and dependence on the difference of the method for solving the SDE 
and the gauge fixing condition.

%%%%%%%%%%%%%%%%%%%%%%%%%%%%%%%%%%%%%%%%%%%%%%%%%%%%%%%%%%%%%%%%%%%%%%%%%%%
%%%%%%%%%%%%%%%%%%%%%%%%%%%%%%%%%%%%%%%%%%%%%%%%%%%%%%%%%%%%%%%%%%%%%%%%%%%
\section{The Schwinger-Dyson equation at Finite $T$}
\label{sec:SDE}
%In this section, we formulate the SDE with the ladder approximation at finite
%$T$.

The SDE for the fermion propagator ${\mathcal S}$ in the imaginary time
formalism is expressed as 
\begin{align}
{\mathcal S}^{-1}(i\omega_n,\vec{p})&-{\mathcal S}_{free}^{-1}(i\omega_n,\vec{p})
=\Sigma(i\omega_n,\vec{\vec{p}})
\notag \\
%\label{eq:ladder-SDE-im} 
&=g^2 T \sum_{m=-\infty}^{\infty}\int\frac{d^3k}{(2\pi)^3}
         \gamma_{\mu}{\mathcal S}(i\omega_m,\vec{k})\gamma_{\nu}
         {\mathcal D^{\mu\nu}}(i\omega_n-i\omega_m,\vec{p}-\vec{k})  ,
\label{eq:ladder-SDE-im} %         \label{eq:spSDE-sigma-im}
\end{align}
where $\Sigma$ and ${\mathcal D}$ are the fermion self-energy and a gauge
boson propagator, respectively, and
$\omega_n=(2n+1)\pi T$ is the Matsubara frequency for the fermion. 
The SDE was employed to describe the spontaneous chiral symmetry breaking in the
strong coupling gauge 
theories\cite{kugo,Harada:1998zq,Harada:2001tr,Takagi,Ikeda:2001vc,Nakkagawa:2007hu}. 
When the value of the gauge coupling $g$ in Eq.~(\ref{eq:ladder-SDE-im}) is
larger than the critical value $g_c$, 
the chiral symmetry is spontaneously broken at low $T$.
When $g$ is smaller than $g_c$, on the other hand,
the chiral symmetry is not broken even at $T=0$.
We summarize the phase structure in appendix~\ref{app:imSDE}.
In the following, we consider only the chiral symmetric phase 
%in Fig.~\ref{fig:tc-g-C} 
to investigate the fermion spectrum.

We start with formulating the SDE on the real time axis following Ref.~\citen{Harada:2008vk}.
The Matsubara summation in Eq.~(\ref{eq:ladder-SDE-im}) can be performed  by
using the spectral representation for fermion and gauge boson propagators
as follows:
\begin{align}
\Sigma(i\omega_n,\vec{p}) 
%         &= g^2 T \sum_{m=-\infty}^{\infty}\int\frac{d^3k}{(2\pi)^3}
%         \gamma_{\mu}\left(-\int dz \frac{\rho_f(z,\vec{k})}{i\omega_m-z}\right)
%         \gamma_{\nu}\left(-\int dz^{\prime}
%         \frac{\rho_B^{\mu\nu}(z^{\prime},\vec{p}-\vec{k})}
%              {i\omega_n-i\omega_m-z^{\prime}}\right)   \\
           = &-g^2 \int\frac{d^3k}{(2\pi)^3} \int dz \int dz^{\prime}
              \frac{1}{i\omega_n-z-z^{\prime}}
              \left(1-f(z)+n(z^{\prime})\right)  
\notag \\
              &\times \left[\gamma_{\mu}\rho_f(z,\vec{k})\gamma_{\nu}
                \rho_B^{\mu\nu}(z^{\prime},\vec{p}-\vec{k})\right],
\end{align}
where $f$ ($n$) is the Fermi-Dirac (Bose-Einstein) distribution
function, and $\rho_f$ and $\rho_B^{\mu\nu}$ are the spectral functions for
the fermion
and the gauge boson introduced as 
\begin{align}
{\mathcal S}(i\omega_m,\vec{p})&=-\int dz \frac{\rho_f(z,\vec{k})}{i\omega_m-z},
\notag\\ 
{\mathcal D}^{\mu\nu}(i\omega_n-i\omega_m,\vec{p}-\vec{k})
&=-\int dz
\frac{\rho^{\mu\nu}_{B}(z^{\prime},\vec{p}-\vec{k})}{i\omega_n-i\omega_m-z^{\prime}
} .
\end{align}
Then we replace the Matsubara frequency as $i\omega_n\rightarrow
p_0+i\epsilon$ to obtain the retarded fermion self-energy as
%Since we are interested in the fermion spectrum, we perform an analytic
%continuation by substituting $i\omega_n\rightarrow p_0+i\epsilon$.
%Thus we obtain the retarded fermion self-energy as follows:
\begin{align}
\Sigma_R(p_0,\vec{p}) = -g^2 \int\frac{d^3k}{(2\pi)^3} \int dz &\int dz^{\prime}
         \frac{1}{p_0+i\epsilon-z-z^{\prime}}
         \left(1-f(z)+n(z^{\prime})\right) 
\notag \\
         &\times\left[\gamma_{\mu}\rho_f(z,\vec{k})\gamma_{\nu}
           \rho_B^{\mu\nu}(z^{\prime},\vec{p}-\vec{k})\right].
\label{eq:spSDE-sigma-re} 
\end{align}
Taking the imaginary parts of both sides,
we have
%$\Sigma_R(p_0,\vec{p})$
%in Eq.~(\ref{eq:spSDE-sigma-re}) is given by 
\begin{align}
{\rm Im }\Sigma_R(p_0,\vec{p}) 
       = \pi g^2 \int^{\Lambda}\frac{d^3k}{(2\pi)^3} 
       &\int_{-\Lambda}^{\Lambda} dk_0 
\left(1-f(k_0)+n(p_0-k_0)\right) 
\notag \\
         &\times \left[\gamma_{\mu}\rho_f(k_0,\vec{k})\gamma_{\nu}
           \rho_B^{\mu\nu}(p_0-k_0,\vec{p}-\vec{k})\right],
         \label{eq:spSDE-Imsigma} 
\end{align}
where $\Lambda$ is the ultraviolet cutoff introduced for regularization.
Using the dispersion relation, the real part of the retarded fermion
self-energy is expressed as
\begin{align}
{\rm Re }\Sigma_R(p_0,\vec{p}) 
      &= \frac{1}{\pi} {\rm P}\int_{-\Lambda}^{\Lambda} dp_0^{\prime}\frac{{\rm Im}
  \Sigma_R(p_0^{\prime},\vec{p})}{p_0^{\prime}-p_0} ,
       \label{eq:spSDE-Resigma}
\end{align}
where ${\rm P}$ denotes the principal integral.
The full retarded fermion propagator is expressed with the retarded fermion 
self-energy $\Sigma_R(p_0,\vec{p})$ as 
\begin{align}
iS_R(p_0,\vec{p})=\frac{-1}{\Sfl{p}+\Sigma_R} 
%  = \frac{-1}{(p_0+\Sigma_0)\gamma_0-(p+\Sigma_{\rm v})
%    \hat{\vec{p}}\cdot\vec{\gamma}}  
.
\label{eq:spSDE-SR}
\end{align}
The spectral function for full fermion propagator 
is in turn expressed with the retarded propagator as
\begin{align}
\rho_f(p_0,p)%=\rho_0\gamma_0-\rho_{\rm v}\hat{\vec{p}}\cdot\vec{\gamma}
             =\frac{1}{\pi}{\rm Im}\left[iS_R(p_0,\vec{p})\right],
\label{eq:spSDE-Sp}
\end{align}
where $p=|\vec{p}|$ and we remark that the fermion spectral function $\rho_f$ has spinor
structure.
Equations (\ref{eq:spSDE-Imsigma}), (\ref{eq:spSDE-Resigma}),
(\ref{eq:spSDE-SR}) and (\ref{eq:spSDE-Sp}) make a closed form 
for the spectral function of fermion $\rho_f$ 
when the spectral function of gauge boson $\rho_B^{\mu\nu}$ is a known function.
In the method used in the previous work\cite{Harada:2007gg}, we 
computed both real and imaginary parts of fermion propagator. 
The present formulation has the advantage of reducing the computation task,
because we only compute the imaginary part of fermion self-energy from the loop
integral, while the real part is computed from the imaginary part using the
dispersion relation.

To make Eqs.~(\ref{eq:spSDE-Imsigma})--(\ref{eq:spSDE-Sp}) a closed form,
we choose an appropriate gauge boson propagator for the system.
In the present work, to study the effects of collective excitations 
of gauge boson, we use the HTL corrected gauge boson propagator in the
Coulomb gauge. 
In the previous work\cite{Harada:2007gg}, we used the Feynman gauge to avoid double pole term.
When we take account of the HTL correction to the gauge boson, 
we encounter the double pole term in the Landau gauge 
as well as in the Feynman gauge. 
In the present work,
for simplicity of computation, we choose the Coulomb gauge in which there is
no double pole term. 
We will discuss the dependence of the fermion spectrum 
on the gauge choice in appendix 
\ref{app:gauge-method-depend} in the case of tree gauge boson propagator.

The spectral function of the HTL corrected gauge boson in the Coulomb gauge is 
expressed as 
\begin{align}
\rho_{B}^{\mu\nu}(q_0,q)= \rho_T(q_0,q)P_{T}^{\mu\nu}+\rho_L(q_0,q)
\delta^{\mu 0}\delta^{\nu 0},
\label{eq:rho-cou}
\end{align}
where $P_T^{\mu\nu}$ is the transverse projection 
$P_T^{00}=P_T^{0i}=0$, and $P_T^{ij}=\delta^{ij}-q^{i}q^{j}/q^2$. 
$\rho_T^{\mu\nu}$ and $\rho_L^{\mu\nu}$ are spectral functions for transverse
and longitudinal parts of the HTL corrected gauge boson propagator. 
They are expressed as \cite{LeBellac}
\begin{align}
\rho_{T}(q_0,q)=Z_T(q)[\delta(q_0-\omega_T(q))-\delta(q_0+\omega_T(q))]
              +\beta_T(q_0,q) ,
\label{eq:rho_T}
\\
\rho_{L}(q_0,q)=Z_L(q)[\delta(q_0-\omega_L(q))-\delta(q_0+\omega_L(q))]
              +\beta_L(q_0,q) ,
\label{eq:rho_L}
\end{align}
where $\omega_T(q)$ and $\omega_L(q)$ are dispersion relations of 
the transverse and longitudinal gauge bosons,
$\beta_T$ and $\beta_L$ are continuum parts of spectral functions and 
$Z_{T}$ and $Z_{L}$ are pole residues.
They are expressed as 
\begin{align}
\beta_T(q_0,q)=\frac{m_g^2 x (1-x^2) \frac{\theta(1-x^2)}{2}}
     {\left(q^2(x^2-1)-m_g^2 \left[
         x^2+\frac{x(1-x)}{2}\ln\left|\frac{x+1}{x-1}\right|\right]\right)^2+\frac{\pi^2m_g^4x^2(1-x^2)^2}{4}} ,
\label{eq:def-beta_T} \\
\beta_L(q_0,q)=\frac{m_g^2 x \theta(1-x^2)}
     {\left[q^2+2m_g^2 \left(
         1-\frac{x}{2}\ln\left|\frac{x+1}{x-1}\right|\right)\right]^2+\pi^2m_g^4x^2} ,
\label{eq:def-beta_L}\\
Z_T(q)=\frac{\omega_T(q)(\omega_T^2(q)-q^2)}{3\omega_p^2\omega_T^2(q)-(\omega_T^2(q)-q^2)},~~ 
Z_L(q)=\frac{\omega_L(q)(\omega_L^2(q)-q^2)}{q^2(q^2+2m_g^2-\omega_L^2(q))} ,
\label{eq:def-res_TL}
\end{align}
where $m_g$ is the thermal mass of gauge boson
and is related with the 
plasma frequency $\omega_p$ as $\omega_p=\sqrt{2/3}m_g=\omega_{T,L}(0)$.
%In QED, the thermal mass of photon is $m=gT/\sqrt{6}$, in this paper, 

It is useful to parameterize  the retarded fermion self-energy $\Sigma_R$ and
the fermion spectral function $\rho_f$ as follows:
\begin{align}
\Sigma_R&=\Sigma_R^0\gamma_0-\Sigma_R^{\rm v} \frac{\vec{p}}{p}\cdot\vec{\gamma},
\notag \\
\rho_f&=\frac{1}{\pi}{\rm Im}~iS_R(p_0,p) 
=\frac{1}{2}\left(\gamma_0-\frac{{\vec{p}}}{p}\cdot\vec{\gamma}\right)\rho_{+}
+
\frac{1}{2}\left(\gamma_0+\frac{{\vec{p}}}{p}\cdot\vec{\gamma}\right)\rho_{-},
\label{eq:spSDE-Sp-pm}
\end{align}
in which $\rho_+$ ($\rho_{-}$) is the fermion spectral function for
(anti-) fermion sector. 
Substituting Eqs.~(\ref{eq:rho-cou}) and (\ref{eq:spSDE-Sp-pm}) into
Eq.~(\ref{eq:spSDE-Imsigma}) 
and making suitable projections, the energy part ($\Sigma_R^0$) and the vector 
part ($\Sigma_R^{\rm v}$) of the retarded self-energy become 
%(Note that $\Sigma_R$ is expressed as 
%$\Sigma_R^0\gamma_0-\Sigma_R^{\rm v} \hat{\vec{p}} \cdot \vec{\gamma}$)
%By using Eq.~(\ref{eq:rho-cou}) and making suitable projection,
%the imaginary part of the retarded fermion self-energy
%Eq.~(\ref{eq:spSDE-Imsigma})  becomes
\begin{align}
{\rm Im }\Sigma_R^{0}(p_0,\vec{p}) 
%       &\equiv
%       \frac{1}{4}{\rm Tr}\left[\gamma_0~{\rm Im}\Sigma_R\right] 
   &= \pi g^2 \int^{\Lambda}\frac{d^3k}{(2\pi)^3} \int^{\Lambda} dk_0 
      \left(1-f(k_0)+n(q_0)\right) 
      \frac{\rho_{+}(k_0,\vec{k})+\rho_{-}(k_0,\vec{k})}{2}
      %\rho_0(k_0,\vec{k}) 
      \notag \\ 
      & \hspace{1em}\times \left[
           2 \rho_T(q_0,q)+\rho_L(q_0,q)
           \right] ,
\label{eq:spSDE-im-sigma0-HTL-p}
\\
%\end{align}
%\begin{align}
{\rm Im }\Sigma_R^{\rm v}(p_0,\vec{p}) 
%&\equiv 
%\frac{1}{4}{\rm Tr}\left[\hat{\vec{p}}\cdot\vec{\gamma}~{\rm
%Im}\Sigma_R\right] \notag \\
&= \pi g^2 \int^{\Lambda}\frac{d^3k}{(2\pi)^3} \int^{\Lambda} dk_0 
\left(1-f(k_0)+n(q_0)\right) 
\frac{\rho_{+}(k_0,\vec{k})-\rho_{-}(k_0,\vec{k})}{2}
\notag \\ 
%\rho_{\rm v}(k_0,\vec{k})
&\hspace{1em}\times 
  \left[
  2 \rho_T(q,q)
  \frac{-pk+(p^2+k^2)\cos\theta-pk\cos^2\theta}{q^2}
  -\rho_L(q_0,q) ~\cos\theta
  \right] , 
\label{eq:spSDE-im-sigmav-HTL-p}
\end{align}
where $q_0=p_0-k_0$ and $q=|\vec{p}-\vec{k}|$.
%$\rho_+$ and $\rho_-$ are fermion spectral functions for fermion sector and
%anti-fermion sector. 
%given in Eq.~(\ref{eq:rho-define}).
%The $\rho_+$ and $\rho_-$ 
%are related with the retarded fermion propagator as 
%\begin{align}
%  \rho_f(p_0,p)=\frac{1}{\pi}{\rm Im}~iS_R(p_0,p) =
%  \frac{\gamma_0-\hat{\vec{p}}\cdot\vec{\gamma}}{2}\rho_{+}
%  + \frac{\gamma_0+\hat{\vec{p}}\cdot\vec{\gamma}}{2}\rho_{-}~.
%\label{eq:spSDE-Sp-pm}
%\end{align}
Equations (\ref{eq:spSDE-im-sigma0-HTL-p}), (\ref{eq:spSDE-im-sigmav-HTL-p}),
(\ref{eq:spSDE-Resigma}), (\ref{eq:spSDE-SR}) and
(\ref{eq:spSDE-Sp-pm}) make a set of iterative equations for the spectral
functions $\rho_{\pm}$ in a closed form.

%%%%%%%%%%%%%%%%%%%%%%%%%%%%%%%%%%%%%%%%%%%%%%%%%%%%%%%%%%%%%%%%%%%%%%%%%%%
%%%%%%%%%%%%%%%%%%%%%%%%%%%%%%%%%%%%%%%%%%%%%%%%%%%%%%%%%%%%%%%%%%%%%%%%%%%
\section{Numerical Results}
\label{sec:NR}
In the following, we fix the cutoff $\Lambda$ as $\Lambda/T=5$,
and we show only $\rho_+$ noting the relation
$\rho_{-}(p_0,p)=\rho_{+}(-p_0,p)$  
obtained from the charge conjugation invariance.

\begin{figure}[t]
  \begin{tabular}{ccc}
    %\hspace{}
    \begin{minipage}{1.0\hsize}
    \begin{minipage}{0.33\hsize}
      \begin{center}
	\includegraphics[keepaspectratio,height=4.3cm]{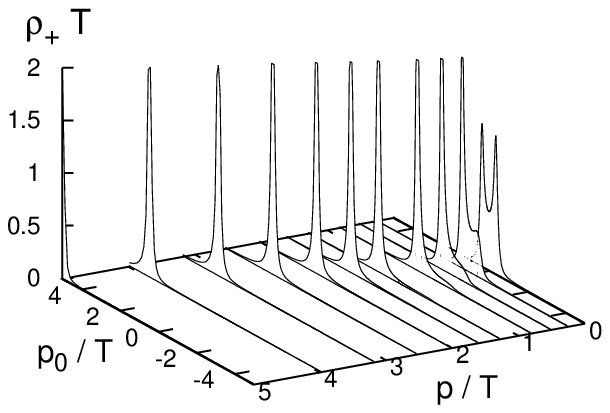}
      \end{center}
    \end{minipage}
    \hspace{-1em}
    \begin{minipage}{0.33\hsize}
      \begin{center}
	\includegraphics[keepaspectratio,height=4.3cm]{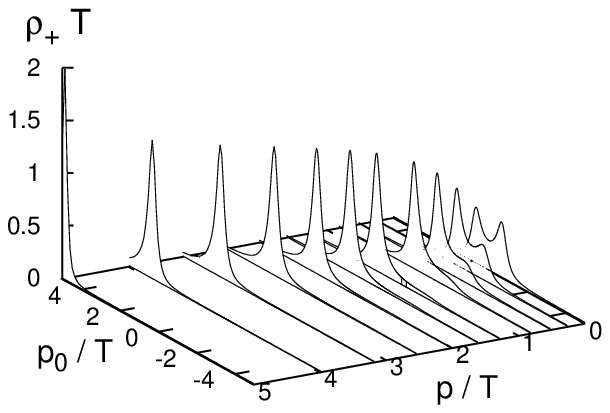}
      \end{center}
    \end{minipage}
    \hspace{-1em}
    \begin{minipage}{0.33\hsize}
      \begin{center}
	\includegraphics[keepaspectratio,height=4.3cm]{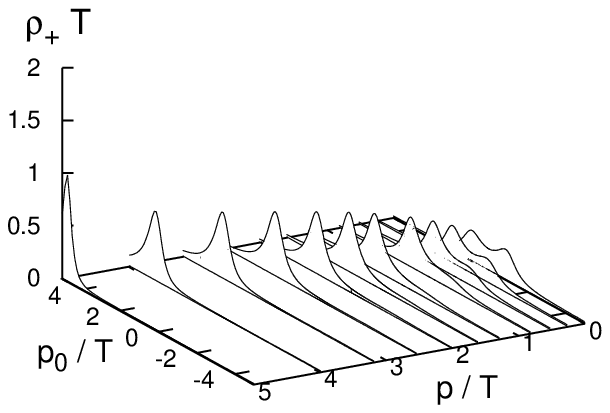}
      \end{center}
    \end{minipage}\\
    \hspace{-2em}
    \begin{minipage}{0.33\hsize}
      \begin{center}
	\includegraphics[keepaspectratio,height=4.3cm]{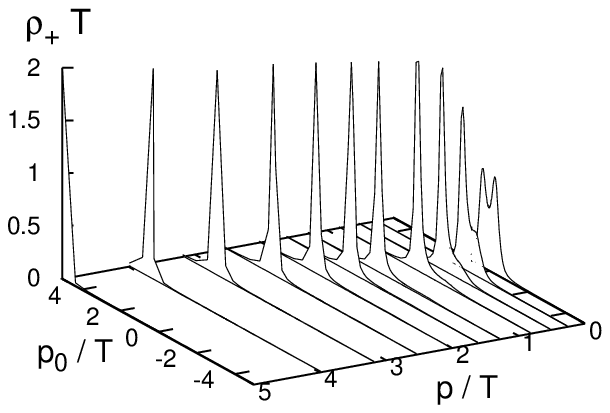}
      \end{center}
    \end{minipage}
    \hspace{-1em}
    \begin{minipage}{0.33\hsize}
      \begin{center}
	\includegraphics[keepaspectratio,height=4.3cm]{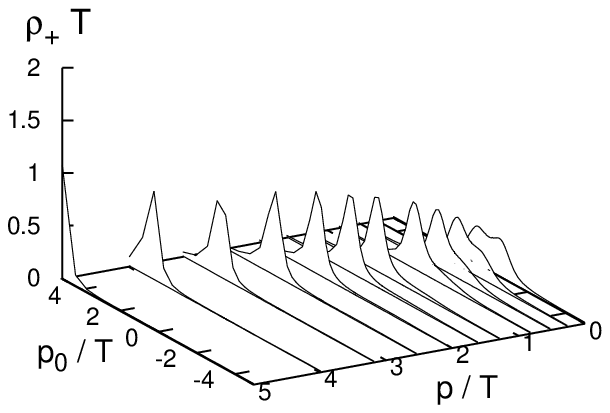}
      \end{center}
    \end{minipage}
    \hspace{-1em}
    \begin{minipage}{0.33\hsize}
      \begin{center}
	\includegraphics[keepaspectratio,height=4.3cm]{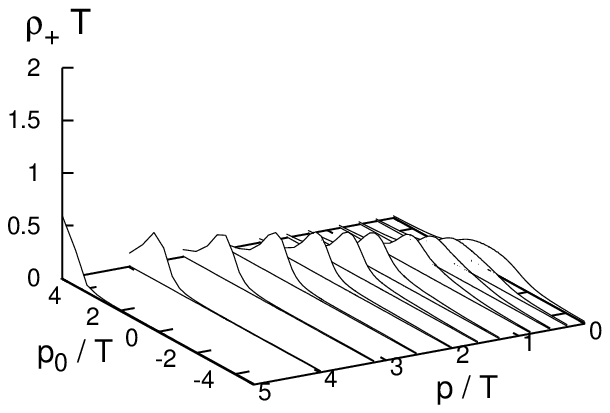}
      \end{center}
    \end{minipage}
    \caption{
      The fermion spectral function $\rho_+$ in the Coulomb gauge.
      Fermion spectra in the upper panels are employing the tree
      gauge boson and those in the lower panels are employing the HTL corrected
      gauge boson 
      with $m_g/T=g/\sqrt{6}$.
      When we go from the left panels to the right panels, the coupling becomes
      stronger 
      in both upper and lower panels: 
      $g=1$ (left panels), $g=2$ (middle panels) and $g=3$ (right panels).
      The figures are clipped at $\rho_+ T = 2$.
    }
    \label{fig:spectral-function-C-m/T}
    \end{minipage}
  \end{tabular}
\end{figure}

In Fig.~\ref{fig:spectral-function-C-m/T}, we plot the fermion spectral
function $\rho_{+}$ for $g=1,~2$ and $3$, where 
three upper panels represent the spectral functions of fermion interacting with
the tree gauge boson and three lower panels represent those of fermion
interacting with the HTL corrected gauge boson with $m_g=gT/\sqrt{6}$.
In the upper panels, 
we see that $\rho_+$ has two peaks in the low momentum region, one in the
positive energy region corresponding to the normal quasi-fermion, and another
in the negative energy region corresponding to the anti-plasmino. 
Two peaks become broad as the coupling increases. 
In the higher momentum region, fermion spectrum approaches the one of free
fermion. 
These results are consistent with those in the previous
work  \cite{Harada:2007gg}. 
Comparing fermion spectra in the lower panels with those in the upper panels,
we find that two peaks in the lower panels are broader than those
in the upper panels, i.e. two peaks become broader by the effects of 
the thermal correction to the gauge boson in medium.
In lower panels, two peaks become broader as the coupling increases as in the
upper panels.
%Disappearance of the anti-plasmino peak in lower panels is
%more rapid than that in upper panels.
At $g=2$, although we can see clear two peaks in the upper panel, 
two peaks in the lower panel are very broad, and 
the quasi-particle picture seems no longer valid.
At $g=3$ in lower panel there is no clear peak, which implies the disappearance 
of quasi-fermions.

The broadening of the peak with increasing coupling, as we stated in the
previous work\cite{Harada:2007gg}, can be understood as the
increasing probability of the 
gauge boson emission and absorption from a fermion. 
Because the SDE contains the effects of multiple scatterings with gauge bosons 
through the self-consistency condition, the peak obtained by the SDE is
broader than the one obtained by the one-loop approximation (see
Ref.~\citen{Harada:2007gg}).

\begin{figure}[t]
%  \begin{tabular}{c}
%      \begin{minipage}{1.0\hsize}
        \begin{center}
          \includegraphics[keepaspectratio,height=6.cm]{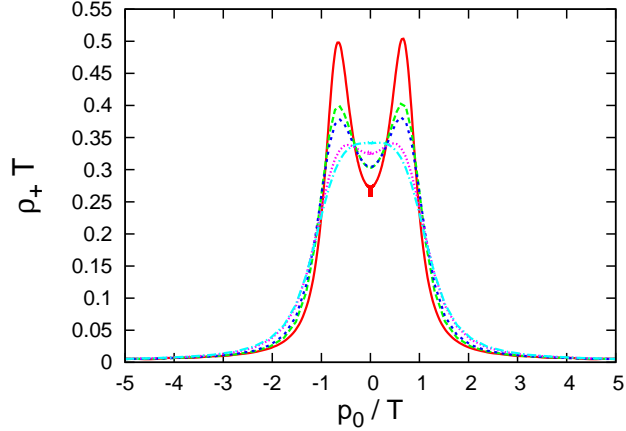}
          \caption{Fermion spectral functions for $g=2$ at $p=0$. 
            The solid (red) curve is for $m_g=0$ (tree), 
            the dashed (green) for  $m_g/T=1/5$, 
            the short-dashed (blue) for  $m_g/T=1/3$,
            the dotted (magenta) for 
            $m_g/T=g /\sqrt{6}$ and 
            the dot-dashed (cyan) for  $m_g/T=1$. 
          } 
          \label{fig:sp-g2-t02}
          \end{center}
%      \end{minipage}
%  \end{tabular}
\end{figure}

The difference between the HTL corrected gauge boson and tree gauge boson lies  
in the difference of the value of $m_g/T$:
The HTL corrected gauge boson becomes tree gauge boson when $m_g/T=0$ in
Eqs.~(\ref{eq:def-beta_T})--(\ref{eq:def-res_TL}). 
Here, we treat $m_g/T$ as a parameter in
Eqs.~(\ref{eq:def-beta_T})--(\ref{eq:def-res_TL}) 
and study $m_g/T$ dependence of the fermion spectrum.  
In Fig.~\ref{fig:sp-g2-t02}, we plot the fermion
spectral functions at $p=0$ for $m_g/T=0$, $1/5$, $1/3$, $g/\sqrt{6}$
and $1$ with $g=2$ fixed.
We see that two peaks become broad as the value of $m_g/T$ increases, and that 
two peaks merge to very broad one peak for $m_g/T=1$.
(See the dot-dashed (cyan) curve in Fig.~\ref{fig:sp-g2-t02})

\begin{figure}[tb]
%  \begin{tabular}{c}
%      \begin{minipage}{1.0\hsize}
        \begin{center}
          \includegraphics[keepaspectratio,height=6.cm]{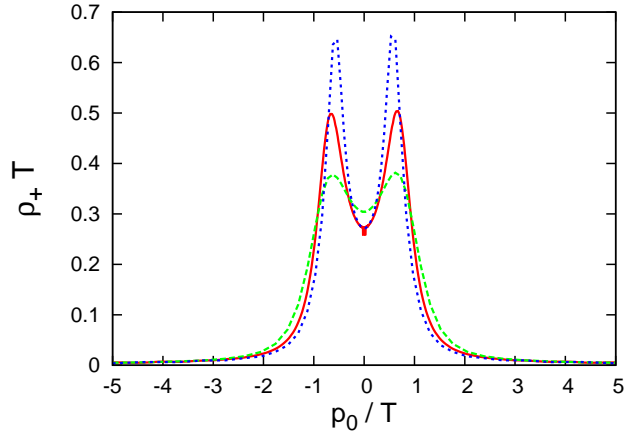}
          \caption{
            Fermion spectral functions for $g=2$ and $m_g/T=1/3$ at $p=0$.  
            Solid (red) curve is the spectrum of fermion coupled with tree
            gauge boson, dashed (green) curve is coupled with ``full HTL'' gauge
            boson expressed in Eqs.~(\ref{eq:rho_T_fullHTL}) and
            (\ref{eq:rho_L_fullHTL}),  
            and dotted (blue) curve is coupled with ``HTL pole'' gauge boson
            expressed in Eqs.~(\ref{eq:rho_T_HTLpole}) and
            (\ref{eq:rho_L_HTLpole}).  
          } 
          \label{fig:sp-part-t02}
        \end{center}
%      \end{minipage}
%  \end{tabular}
\end{figure}

To understand the broadening of the peak with increasing value of $m_g/T$, 
we consider the case in which the fermion interacts with HTL corrected
gauge boson without continuum parts $\beta_T$ and $\beta_L$ 
in Eqs.~(\ref{eq:rho_T}) and (\ref{eq:rho_L}).  
We call this gauge boson ``HTL pole'' gauge boson: 
\begin{align}
\rho_{T}(q_0,q)\big|_{\rm ``HTL~pole"}=Z_T(q)[\delta(q_0-\omega_T(q))-\delta(q_0+\omega_T(q))]~~,
\label{eq:rho_T_HTLpole}
\\
\rho_{L}(q_0,q)\big|_{\rm ``HTL~pole"}=Z_L(q)[\delta(q_0-\omega_L(q))-\delta(q_0+\omega_L(q))]~~.
\label{eq:rho_L_HTLpole}
\end{align}
On the other hand, the HTL corrected gauge boson with 
continuum part is labeled as ``full HTL'' gauge boson:
\begin{align}
\rho_{T}(q_0,q)\big|_{\rm
  ``full~HTL"}=Z_T(q)[\delta(q_0-\omega_T(q))-\delta(q_0+\omega_T(q))]
  + \beta_{T}(q_0,q)~~,
\label{eq:rho_T_fullHTL} \\
\rho_{L}(q_0,q)\big|_{\rm
  ``full~HTL"}=Z_L(q)[\delta(q_0-\omega_L(q))-\delta(q_0+\omega_L(q))]
   + \beta_{L}(q_0,q)~~. 
\label{eq:rho_L_fullHTL}
\end{align}
We plot the spectral functions of fermion at $p=0$ coupled with tree gauge
boson, the ``full HTL'' gauge boson and the ``HTL pole'' gauge boson in
Fig.~\ref{fig:sp-part-t02}. 

The spectral function of fermion interacting with the ``HTL pole'' gauge boson
shown by the dotted (blue) curve in Fig.~\ref{fig:sp-part-t02} is sharper than
that with tree gauge boson shown by the solid (red) curve.
Because the SDE is a self-consistent equation, the internal fermion propagator 
has a thermal mass. 
The imaginary part of the fermion self-energy at one-loop, in which 
both the internal fermion and gauge boson have pole masses (denoted by $m_f$
and $m_g$), is forbidden in the energy region $|m_f-m_g|<|p_0|<m_f+m_g$ at
$p=0$ owing to the energy-momentum conservation.
While the imaginary part in the SDE is not completely forbidden but 
suppressed in the region,
since the internal fermion in the SDE has a width.
%Because the SDE is a self-consistent equation, the internal fermion propagator 
%has a thermal mass. 
%%When a fermion and a gauge boson have masses $m_f$ and $m_g$ respectively
%%and they exist as poles with no widths, 
%Let us consider a one-loop correction to the fermion self-energy, in which 
%both the internal fermion and gauge boson have pole masses, denoted by $m_f$
%and $m_g$, respectively.
%Then
%the imaginary part of the fermion self-energy at one-loop is forbidden in the energy
%region $|m_f-m_g|<|p_0|<m_f+m_g$ at $p=0$ owing to the energy-momentum
%conservation.
%Because the internal fermion in the SDE has a width, the imaginary part is not 
%completely forbidden, but it is suppressed in the region
%$|m_f-m_g|<|p_0|<m_f+m_g$.
As was observed in a Yukawa model in Ref.~\citen{Mitsutani:2007gf},
the suppression of the imaginary part leads to a sharp peak in the spectral
function. 
As a result, the peak of 
the fermion spectrum interacting with the ``HTL pole'' gauge boson
is sharper than that with the tree gauge boson.
As $m_g$ increases, the above region becomes larger and the peak of fermion
spectrum becomes even sharper.
When the value of $m_g$ is very large,  the fermion spectrum will approach
that of the free fermion because the gauge boson decouples from
the system.

\begin{figure}[t]
  \begin{center}
    \includegraphics[keepaspectratio,height=3cm]{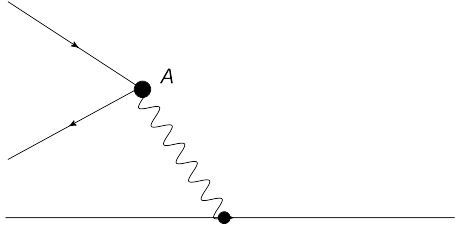}
    \caption{A diagram which generates the imaginary part of the propagator
      for the fermion interacting with the HTL corrected gauge boson.}  
    \label{fig:Landau}
  \end{center}
\end{figure}

From the above analysis, we find that the continuum parts in
Eqs.~(\ref{eq:rho_T_fullHTL}) and 
(\ref{eq:rho_L_fullHTL}) play important rolls for the broadening of the
fermion spectrum in the case of the ``full HTL'' gauge boson. 
Below, we give an interpretation for the broadening of the fermion spectrum
with increasing $m_g/T$:
The dominant parts of $\beta_T$ and $\beta_L$ are proportional to $m_g^2$ 
as we see in Eqs.~(\ref{eq:def-beta_T}) and (\ref{eq:def-beta_L}), 
i.e. the Landau damping of gauge boson becomes larger as the value of 
$m_g/T$ increases.
$m_g/T$ is proportional to 
the coupling at the point indicated by ``A'' in
Fig.~\ref{fig:Landau}.  
[Note that the coupling at ``A'' is different from $g$.]
Therefore the broadening can be interpreted as the increasing probability 
of scatterings of fermion and gauge boson at ``A''.
We would like to note that this effects is not included in quenched lattice
simulations.

\begin{figure}[t]
%  \begin{tabular}{c}
%      \begin{minipage}{1.0\hsize}
        \begin{center}
          \includegraphics[keepaspectratio,height=6.cm]{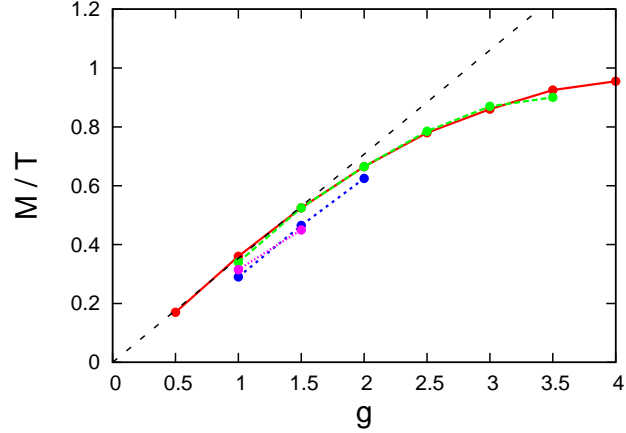}
          \caption{
            The coupling dependence of the thermal mass $M/T$. 
            The solid (red) curve represents the one for $m_g=0$ (tree),
            the dashed (green) curve for $m_g/T=1/3$, 
            the short-dashed (blue) curve for $m_g/T=1/2$, 
            the dotted (magenta) curve for $m_g/T=1/\sqrt{6}$
            and the dashed (black) line represents the thermal
            mass of fermion in the HTL approximation. 
            In the case of $m_g/T=1$, we cannot read the peak
            as an excitation.}
          \label{fig:M-g-t02-0uv0}
        \end{center}
%      \end{minipage}
%  \end{tabular}
\end{figure}

\begin{figure}[t]
%  \begin{tabular}{c}
%      \begin{minipage}{1.0\hsize}
        \begin{center}
          \includegraphics[keepaspectratio,height=6.cm]{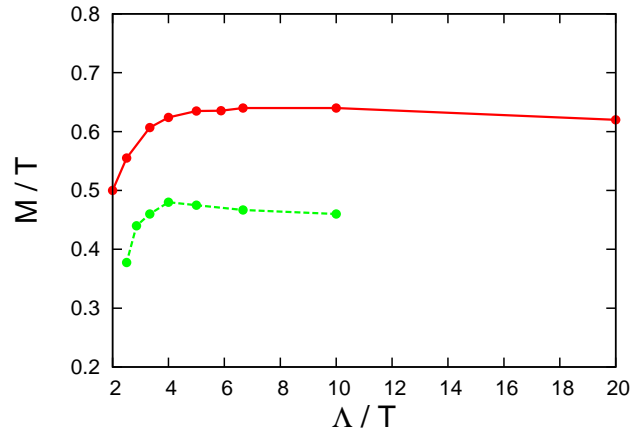}
          \caption{Cutoff dependence of the thermal mass $M/T$. 
            The solid (red) curve is for $m_g/T=1/3$ and  
            the dashed (green) for $m_g/T=g /\sqrt{6}$.
            For $\Lambda/T\gtrsim 4$, the ratio $M/T$ is almost independent of
            the cutoff.}  
          \label{fig:M-T-g2}
        \end{center}
%      \end{minipage}
%  \end{tabular}
\end{figure}

At the last of this section, 
we consider the coupling dependence (Fig.~\ref{fig:M-g-t02-0uv0}) and the
cutoff dependence (Fig.~\ref{fig:M-T-g2}) of the peak 
position of the fermion spectral function which we call the thermal mass.
Figure~\ref{fig:M-g-t02-0uv0} shows that the thermal masses hardly depend on
the value of $m_g/T$ when the peaks exist as clear ones. 
On the other hand, the widths strongly depend on $m_g/T$ as seen in
Fig.~\ref{fig:sp-g2-t02},
and there are no quasi-fermions in the strong coupling region as seen in 
Fig.~\ref{fig:spectral-function-C-m/T}.
That is why the curves for $m_g/T>0$ in Fig.~\ref{fig:M-g-t02-0uv0} 
terminate at some values of the coupling.
This is contrasted to  
the previous work\cite{Harada:2007gg}, in which tree gauge boson ($m_g/T=0$)
was used.  
There, we saw that the peak position saturates at a value of coupling and the
peak position $M$ 
becomes $M/T\sim 1$ in the very strong coupling region. 
{}From Fig.~\ref{fig:M-T-g2},
we see that $M/T$ depends little on
$\Lambda/T$ in the region $\Lambda/T \gtrsim 4$. 
We have also checked that the shape of the fermion spectral function 
around the peak hardly depends on the cutoff.

%
%we check the cutoff $\Lambda$ dependence of the thermal mass of
%fermion $M/T$. (See Fig.~\ref{fig:M-T-g2}.)
%We see that $M/T$ depends little on
%$\Lambda/T$ in the region $\Lambda/T \gtrsim 4$. 
%We have also checked that the shape of the fermion spectral function 
%around the peak hardly depends on the cutoff.

%%%%%%%%%%%%%%%%%%%%%%%%%%%%%%%%%%%%%%%%%%%%%%%%%%%%%%%%%%%%%%%%%%%%%%%%%%%
%%%%%%%%%%%%%%%%%%%%%%%%%%%%%%%%%%%%%%%%%%%%%%%%%%%%%%%%%%%%%%%%%%%%%%%%%%%
\section{A summary and discussions}
\label{sec:summary}
In this paper, we investigated the fermion spectra in the chiral
symmetric phase, focusing on the effect of medium correction for gauge boson, 
by employing the Schwinger-Dyson
equation (SDE) for fermion with the ladder approximation.

We found that in-medium effects for gauge boson propagator 
make the peak of the fermion spectral function broader 
compared with the previous result\cite{Harada:2007gg} obtained with the tree 
gauge boson propagator. 
The broadening occurs owing to multiple %increasing probability of
fermion-fermion scatterings in addition to fermion-gauge boson
scatterings, included through the Landau damping of in-medium corrected gauge
boson. 
Thus, it is important to take into account of not only the
thermal mass but also the Landau damping of gauge boson propagator.
Our results show no clear peak in the strong  coupling region, which implies
the disappearance of quasi-fermions in the strongly coupled
plasma.

We would like to make a comment  on the Van Hove singularities.
They exist
in the case of the self-energy consisting of
two very sharp particles whose dispersion relations are different from each
other as in the fermion and gauge boson in the
HTL approximation.
However, if the particles have widths, these singularities are weakened. 
In the SDE, because the (internal) fermion has a broad width 
by non-perturbative effects through the self-consistency condition, we cannot
see the effects of Van Hove singularities in the present analysis.

%%%%%%%%%%%%%%%%%%%
%%%%%%%%%%%%%%%%%%%
In this paper, 
we solved the SDE in the Minkowski space as an iterative equation for the 
fermion spectral function.
In the previous work\cite{Harada:2007gg}, on the other hand, the solution is
obtained by performing the analytic continuation from the imaginary time axis
to the real time axis through an integral equation.
There is little difference in the fermion spectra obtained by two methods
as seen in appendix~\ref{app:gauge-method-depend}.
We also show that there is little gauge dependence of the peak position 
in appendix~\ref{app:gauge-method-depend}.

We discuss the application of the results of the present work to QCD. 
In the very high temperature region of QCD, 
the HTL approximation is valid for studying the quark spectrum.
In the HTL approximation, the fixed coupling, which is the
running coupling at an energy scale of order $T$, 
say $g(T)$, is used, because the quarks and gluons with energy $T$ give
the dominant contribution.
In studying the quark spectrum in the lower $T$ region,
the diagrams other than the HTL start to give non-negligible corrections, 
since the (fixed) coupling $g(T)$ is larger.
For this reason, we can state that, in our approach, 
we include a part of non-perturbative corrections
from a certain diagrams by solving the Schwinger-Dyson equation.
However, the following points of QCD are not included:
(1) the asymptotic freedom through the running effects of the gauge coupling, 
(2) vertex corrections and
(3) higher oder corrections for the gauge boson.

First, we discuss the effect of asymptotic freedom. 
In our analysis, the fixed coupling is used,
which can be understood as 
the running coupling at an energy scale of oder $T$, i.e. $g(T)$.
In the energy scale $E$ lower than $T$,  
the gauge coupling does not run, and we have $g(E)\sim g(T)$. 
This is because quarks and gluons decouple in this energy region owing to the
thermal mass.  
In $E \gtrsim T$, gauge coupling is running, 
and we have $g(E)<g(T)$.
The difference between $g(E)$ and $g(T)$ might cause different results when we 
use the running coupling in the SDE analysis.
However, in the present analysis, we have seen that peak position only depend
on $T$ which is an infrared scale, but does not depend on the cutoff $\Lambda$
which is an ultraviolet scale: The peak position is determined by the
dynamics of the infrared scale only. 
Therefore, we think that the peak positions at least will hardly change from
those in the present analysis.

Second, we discuss vertex corrections. 
When we take account of some vertex corrections, 
the peak of fermion spectrum will become even broader than that in the present
analysis owing to multiple scatterings. 
Then the following qualitative structure of the quark spectrum will hardly
change: 
The peak position of the quark is proportional to the coupling for small $g(T)$
and the quasi-particle picture will be even worse.

Third, we discuss higher order corrections for the gauge boson propagator.
There is a coupled SDE for fermion and gauge boson,  
in which we can include certain higher order corrections for the gauge boson
propagator. 
The peak of fermion spectrum will become even broader than
that in the present analysis owing to multiple scatterings. 
The peak position does not depend on a detailed structure of gauge boson
propagator in the present analysis,
which indicates that the peak position does not change from the present value
in the small coupling region, and that the quasi-particle picture will not be 
good in the strong coupling region.

To summarize, the following qualitative structure of the quark
spectrum will hold in hot QCD: 
In the high $T$ region where $g(T)$ is small, the thermal masses of the
quasi-quark and the plasmino are proportional to the coupling. 
In the low $T$ region where $g(T)$ is large, the width will become broad due
to multiple scatterings. 
Especially, in the strongly coupled QGP near $T_c$, the quasi-particle picture
may be no longer valid.

It is suggested that bound states for light quarks exist near $T_c$
in Ref.~\citen{Shuryak:2004cy} and \citen{Brown:2003km}.
It is also suggested that new mesonic states consisting of a quark and a
plasmino exist in Ref.~\citen{Weldon:1998ic}. 
It is very interesting to study these bound states for light quarks 
through the Bethe-Salpeter equation 
using the fermion spectra obtained in the present analysis.

 %%%%%%%%%%%%%%%%%%%%%%%%%%%%%%%%%%%%%%%%%%%%%%%%%%%%%%%%%%%%%%%%%%%%
%\section*{Acknowledgements}
%We would like to thank ...........
\section*{Acknowledgments} 
This work is supported
in part by the JSPS Grant-in-Aid for Scientific Research
(c) 20540262 and Global COE Program ``Quest for
Fundamental Principles in the Universe'' of Nagoya
University provided by Japan Society for the Promotion
of Science (G07).

 %%%%%%%%%%%%%%%%%%%%%%%%%%%%%%%%%%%%%%%%%%%%%%%%%%%%%%%%%%%%%%%%%%%%
%\appendix
%\section{First Appendix} %Empty argument \section{} yields `Appendix'. 
%
%\section{Second Appendix}
\appendix
\section{The SDE in the imaginary time formalism and the chiral phase transition}
\label{app:imSDE}
%% Coulomb gauge
In this appendix, we determine the critical temperature $T_c$ of the chiral
phase transition from the ladder SDE.
We solve the SDE  in the imaginary time formalism
because the numerical cost is much smaller than solving the SDE on the
real time axis. 

The ladder SDE in the imaginary time formalism is given by
Eq.~(\ref{eq:ladder-SDE-im}).
The full fermion propagator in the imaginary time formalism is restricted 
by the rotational invariance and the parity invariance as follows:
\begin{align}
{\mathcal  S}(i\omega_n,\vec{p})
=\frac{1}{B(i\omega_n,p)+A(i\omega_n,p)\vec{p}\cdot\vec{\gamma}
  -C(i\omega_n,p)i\omega_n\gamma_0} .
\end{align}
The HTL corrected gauge boson propagator in the Coulomb gauge is expressed as
\begin{align}
{\mathcal D}^{\mu\nu}(i\omega_n-i\omega_m,\vec{p}-\vec{k})
&=\Delta_T(i\omega_n-i\omega_m,\vec{p}-\vec{k})P^{\mu\nu}_T
+\Delta_L(i\omega_n-i\omega_m,\vec{p}-\vec{k})\delta^{\mu 0}\delta^{\nu 0},
\label{eq:de-coulomb-prop}
\end{align}
where $\Delta_T$ and $\Delta_L$ are transverse and longitudinal 
HTL corrected gauge boson propagators.
They are expressed as 
\begin{align}
\Delta_T(i\omega,q)
&=\frac{1}{\omega^2+q^2+m_g^2
  \Biggl(
  -\frac{\omega^2}{q^2}
  +\left(1+\frac{\omega^2}{q^2}\right)\frac{|\omega|}{2 q}
  \left( \pi - 2 \tan^{-1}\frac{|\omega|}{q}\right)
  \Biggr)} ,
\label{eq:E-delta_t}
\\
\Delta_L(i\omega,q)
&=\frac{-1}{q^2 + 2 m_g^2
  \Biggl(
   1 - \frac{|\omega|}{2 q}
  \left( \pi - 2 \tan^{-1}\frac{|\omega|}{q}\right)
  \Biggr)} .
\label{eq:E-delta_l}
\end{align}
Note that Eq. (\ref{eq:E-delta_t}) and Eq. (\ref{eq:E-delta_l}) have no 
poles for real $\omega$.
Using these fermion propagator and gauge boson propagators, 
the ladder SDE~(\ref{eq:ladder-SDE-im}) is decomposed into three coupled
equations for $B$, $A$ and $C$ as  
\begin{align}
B&= g^2 T\sum_{m=-\infty}^{\infty}  \int\frac{d^3k}{(2\pi)^3}
   \frac{B}{B^2+C^2 \omega_m^2 + A^2 k^2} 
   \left[
   2 \Delta_T - \Delta_L
   \right],
\label{eq:B-Coulomb-HTL}
\\
A&= 1+ \frac{g^2 T}{p^2} \sum_{m=-\infty}^{\infty}  \int\frac{d^3k}{(2\pi)^3}
   \frac{A}{B^2+C^2 \omega_m^2 + A^2 k^2}
   \left[
   2 \Delta_T 
   \left( \frac{(\vec{p}\cdot\vec{q})(\vec{k}\cdot\vec{q})}{q^2}
   \right)
   - \Delta_L \vec{p}\cdot\vec{k} 
   \right],
\label{eq:A-Coulomb-HTL}
\\
C&= 1+ \frac{g^2 T}{p_0} \sum_{m=-\infty}^{\infty}  \int\frac{d^3k}{(2\pi)^3}
   \frac{C k_0}{B^2+C^2 \omega_m^2 + A^2 k^2} 
   \left[
   2\Delta_T + \Delta_L
   \right].
\label{eq:C-Coulomb-HTL}
\end{align}
Because the integrals in
Eqs.~(\ref{eq:B-Coulomb-HTL})--(\ref{eq:C-Coulomb-HTL}) are divergent, 
we introduce the three-dimensional ultraviolet cutoff $\Lambda$. 
We also truncate the infinite sum of the Matsubara frequency
at a finite number $N$.
However, $N$ can be taken to be sufficiently large so that the 
following results do not depend on it. 
The angle integral in Eqs.~(\ref{eq:B-Coulomb-HTL})--(\ref{eq:C-Coulomb-HTL}) 
are numerically performed.

\begin{figure}[t]
\begin{center}
  \includegraphics[keepaspectratio,height=6cm]{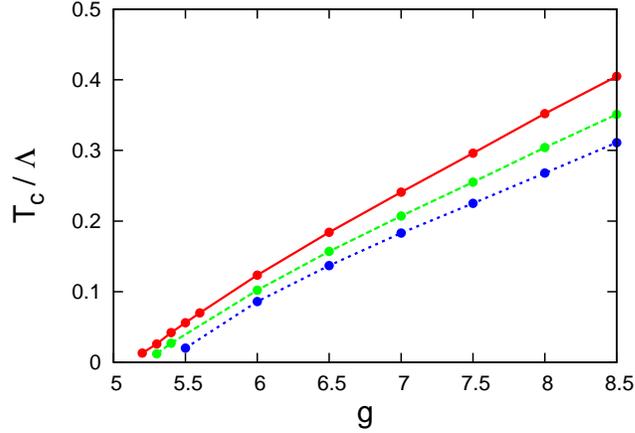}
     \caption{Coupling dependence of the critical temperature
        $T_c/\Lambda$.
	The chiral symmetric (broken) phase is above (below) the line. 
        The solid (red) line shows $T_c/\Lambda$ for $m/\Lambda=0$, 
        the dashed (green) line for $m/\Lambda=0.1$ and the dotted (blue) line
        for $m_g/\Lambda=0.2$.
     }
      \label{fig:tc-g-C}
\end{center}
\end{figure}

In Fig.~\ref{fig:tc-g-C}, we plot the coupling dependence of $T_c/\Lambda$
for $m_g/\Lambda=0,~0.1$ and $0.2$. 
We see that $T_c/\Lambda$ decreases as $m_g/\Lambda$ increases owing to the
screening effects.
This is consistent with the results in Ref.~\citen{Takagi} which 
show that $T_c$ becomes small by the effects of the Debye screening.

%%%%%%%%%%%%%%%%%%%%%%%%%%%%%%%%%%%%%%%%%%%%%%%%%%%%%%%%%%%%%%%%%%%%%%%%%%
\section{Dependences on the gauge fixing condition and  method for solving the
  SDE} 
\label{app:gauge-method-depend}
\begin{figure}[!ht]
  \begin{center}
    \includegraphics[keepaspectratio,height=6cm]{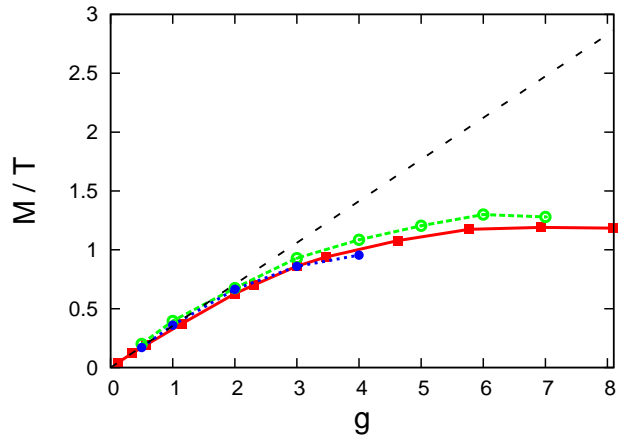}
    \caption{
      Coupling dependence of the thermal mass.
      The solid (red) curve represent the thermal mass obtained by the
      method 1 in the Feynman gauge.
      The dashed (green) curve and dotted (blue) curve represent the thermal
      mass obtained by the method 2, in the Feynman and Coulomb gauge,
      respectively.
      The dashed (black) line represents the thermal mass in the HTL
      approximation.} 
    \label{M-g-tree-gauge-method}
  \end{center}
\end{figure}

There are two methods to obtain the solutions of the SDE on the real time
axis.
One is to perform the analytic continuation for the solutions of the SDE
in the imaginary time formalism through an integral equation\cite{ac}, 
which was used in the previous work\cite{Harada:2007gg}.
Another method is to solve a self-consistent equation as an integral equation
for fermion spectral function\cite{Harada:2008vk}.
In this appendix, we show that the dependence 
of the peak position of the fermion spectral functions on two methods
together with the gauge dependence of the peak position. 
In Fig.~\ref{M-g-tree-gauge-method}, we plot the 
peak positions of fermion spectral functions 
obtained by the method~1 in the Feynman gauge, by the method~2 in the Feynman
gauge and by the method~2 in the Coulomb gauge. 
We find that the peak positions little depend on neither the method nor 
the gauge fixing.
This is consistent with the result of Ref.~\citen{Baym:1992eu}
in which the exact calculation of the fermion self-energy at one-loop level is
performed and it is shown that the peak positions little depend on the gauge
choice.

\begin{thebibliography}{99}
%%%%%%%%%%%%%%%%%%%%%%%%%%%%%%%%%%%%%%%%%%%%%%%%%%%%%%%%%%%%%
% Some macros are available for the bibliography:
%  o for general use
%    \JL : general journals                 \andvol : Vol (Year) Page
%  o for individual journal 
%    \AJ   : Astrophys. J.           \NC         : Nuovo Cim.
%    \ANN  : Ann. of Phys.           \NPA, \NPB  : Nucl. Phys. [A,B]
%    \CMP  : Commun. Math. Phys.     \PLA, \PLB  : Phys. Lett. [A,B]
%    \IJMP : Int. J. Mod. Phys.      \PRA - \PRE : Phys. Rev. [A-E]     
%    \JHEP : J. High Energy Phys.    \PRL        : Phys. Rev. Lett.
%    \JMP  : J. Math. Phys.          \PRP        : Phys. Rep.
%    \JP   : J. of Phys.             \PTP        : Prog. Theor. Phys.     
%    \JPSJ : J. Phys. Soc. Jpn.      \PTPS       : Prog. Theor. Phys. Suppl.
% Usage:
%  \PRD{45,1990,345}          ==> Phys.~Rev.\ \textbf{D45} (1990), 345
%  \JL{Nature,418,2002,123}   ==> Nature \textbf{418} (2002), 123
%  \andvol{B123,1995,1020}    ==> \textbf{B123} (1995), 1020
%%%%%%%%%%%%%%%%%%%%%%%%%%%%%%%%%%%%%%%%%%%%%%%%%%%%%%%%%%%%%
  
%\bibitem{}
\bibitem{Weldon:1982bn}
%\bibitem{Klimov:1981ka}
  V.~V.~Klimov,
  Sov.\ J.\ Nucl.\ Phys.\  {\bf 33}  (1981), 934;
  Yadern.\ Fiz.\  {\bf 33}  (1981), 1734.
  %%CITATION = YAFIA,33,1734;%%

%
%\cite{Weldon:1982aq}
%\bibitem{Weldon:1982aq}
  H.~A.~Weldon,
  %``Covariant Calculations At Finite Temperature: The Relativistic Plasma,''
  Phys.\ Rev.\  D {\bf 26} (1982), 1394.
  %%CITATION = PHRVA,D26,1394;%%.

  H.~A.~Weldon,
  %``Effective Fermion Masses Of Order Gt In High Temperature Gauge Theories
  %With Exact Chiral Invariance,''
  Phys.\ Rev.\  D {\bf 26} (1982), 2789.
  %%CITATION = PHRVA,D26,2789;%%

%
%\cite{Weldon:1989ys}
%\bibitem{Weldon:1989ys}
  H.~A.~Weldon,
  %``DYNAMICAL HOLES IN THE QUARK - GLUON PLASMA,''
  Phys.\ Rev.\  D {\bf 40} (1989), 2410.
  %%CITATION = PHRVA,D40,2410;%%
%
\bibitem{LeBellac}
  M. Le Bellac, 
  {\it Thermal Field Theory 
  (Cambridge University press, Cambridge, England, 1996).}
%


\bibitem{Arsene:2004fa}
  I.~Arsene {\it et al.},
  Nucl.\ Phys.\ A {\bf  757} (2005), 1.
  %[arXiv:nucl-ex/0410020].
  %%CITATION = NUPHA,A757,1;%%

  B.~B.~Back {\it et al.},
  Nucl.\ Phys.\ A {\bf  757} (2005), 28.
  %[arXiv:nucl-ex/0410022].
  %%CITATION = NUPHA,A757,28;%%

  J.~Adams {\it et al.},
  Nucl.\ Phys.\ A {\bf  757} (2005), 102.
  %[arXiv:nucl-ex/0501009].
  %%CITATION = NUPHA,A757,102;%%

  K.~Adcox {\it et al.},
  Nucl.\ Phys.\ A {\bf  757} (2005), 184.
  %[arXiv:nucl-ex/0410003].
  %%CITATION = NUPHA,A757,184;%%


\bibitem{Asakawa:2003re}
  T.~Umeda, K.~Nomura and H.~Matsufuru,
  Eur.\ Phys.\ J.\ C {\bf 39S1} (2005), 9
  [arXiv:hep-lat/0211003].
  %%CITATION = EPHJA,C39S1,9;%%

  M.~Asakawa and T.~Hatsuda,
  Phys.\ Rev.\ Lett.\  {\bf 92} (2004), 012001
  [arXiv:hep-lat/0308034].
  %%CITATION = PRLTA,92,012001;%%

  S.~Datta, F.~Karsch, P.~Petreczky and I.~Wetzorke,
  Phys.\ Rev.\ D  {\bf  69}  (2004), 094507
  [arXiv:hep-lat/0312037].
  %%CITATION = PHRVA,D69,094507;%%

  H.~Iida, T.~Doi, N.~Ishii, H.~Suganuma and K.~Tsumura,
  %``Charmonium properties in deconfinement phase in anisotropic lattice  QCD,''
  Phys.\ Rev.\  D {\bf 74}  (2006), 074502.
  %[arXiv:hep-lat/0602008].
  %%CITATION = PHRVA,D74,074502;%%
%
  A.~Jakovac, P.~Petreczky, K.~Petrov and A.~Velytsky,
  Phys.\ Rev.\  D {\bf 75}  (2007), 014506.
  %[arXiv:hep-lat/0611017].
  %%CITATION = PHRVA,D75,014506;%%

  G.~Aarts, C.~Allton, M.~B.~Oktay, M.~Peardon and J.~I.~Skullerud,
  %``Charmonium at high temperature in two-flavor QCD,''
  Phys.\ Rev.\  D {\bf 76}  (2007), 094513
  %[arXiv:0705.2198 [hep-lat]].
  %%CITATION = PHRVA,D76,094513;%%

\bibitem{Schaefer:1998wd}
  A.~Schaefer and M.~H.~Thoma,
  Phys.\ Lett.\ B {\bf  451}  (1999), 195.

  A.~Peshier and M.~H.~Thoma, 
  Phys.\ Rev.\ Lett.\  {\bf 84}  (2000), 841.

\bibitem{Mannarelli:2005pz}
  M.~Mannarelli and R.~Rapp,
  Phys.\ Rev.\ C  {\bf  72}  (2005), 064905.
  %[arXiv:hep-ph/0505080].
  %%CITATION = PHRVA,C72,064905;%%

\bibitem{Kitazawa:2005mp}
  M.~Kitazawa, T.~Kunihiro and Y.~Nemoto,
  Phys.\ Lett.\ B {\bf  633}  (2006), 269.
  %[arXiv:hep-ph/0510167].
  %%CITATION = PHLTA,B633,269;%%


%\cite{Harada:2007da}
%\bibitem{Harada:2007da}
\bibitem{Harada:2007gg}
  M.~Harada, Y.~Nemoto and S.~Yoshimoto,
  %``Quark spectrum above the critical temperature from Schwinger-Dyson
  %equation,''
   Int.\ J.\ Mod.\ Phys.\ E {\bf 16} (2007), 2282
  [arXiv:hep-ph/0702253].
  %%CITATION = HEP-PH/0702253;%%
%
%\cite{Harada:2007gg}
%\bibitem{Harada:2007gg}
  M.~Harada, Y.~Nemoto and S.~Yoshimoto,
  %``Quasi-quark spectrum in the chiral symmetric phase from the Schwinger-Dyson
  %equation,''
  Prog.\ Theor.\  Phys {\bf 119} (2008), 117.
  arXiv:0708.3351 [hep-ph].
  %%CITATION = ARXIV:0708.3351;%%

\bibitem{Harada:2008vk}
  M.~Harada and Y.~Nemoto,
  %``Quasi-fermion spectrum at finite temperature from coupled Schwinger-Dyson
  %equations for a fermion-boson system,''
  Phys.\ Rev.\  D {\bf 78} (2008) 014004.
  %[arXiv:0803.3257 [hep-ph]].
  %%CITATION = PHRVA,D78,014004;%%
%

\bibitem{Karsch:2007wc}
  F.~Karsch and M.~Kitazawa,
  %``Spectral properties of quarks above T_c in quenched lattice QCD,''
  Phys.\ Lett.\  B {\bf 658}  (2007), 45. 
  %[arXiv:0708.0299 [hep-lat]].
  %%CITATION = PHLTA,B658,45;%%

\bibitem{kugo}
  See, e.g.,
  T. Kugo, 
  in {\it Proc. of 1991 Nagoya Spring School on Dynamical Symmetry 
  Breaking, Nakatsugawa, Japan, 1991,} ed. K. Yamawaki (World Scientific,
  Singapore, 1992).

  V. A. Miransky, 
  {\it Dynamical symmetry breaking in quantum field theories}
  (Singapore, Singapore: World Scientific, 1993).

\bibitem{Harada:1998zq}
  M.~Harada and A.~Shibata,
  Phys.\ Rev.\  D {\bf 59}  (1999), 014010. 
  %[arXiv:hep-ph/9807408].
  %%CITATION = PHRVA,D59,014010;%%

  C.~D.~Roberts and S.~M.~Schmidt,
  Prog.\ Part.\ Nucl.\ Phys.\  {\bf 45}  (2000), S1.
  %[arXiv:nucl-th/0005064].
  %%CITATION = PPNPD,45,S1;%%

%\cite{Harada:2001tr}
\bibitem{Harada:2001tr}
  M.~Harada and S.~Takagi,
  %``Phase transition in two-flavor dense QCD from the Schwinger-Dyson
  %equation,''
  Prog.\ Theor.\ Phys.\  {\bf 107} (2002), 561.
  %[arXiv:hep-ph/0108173]
  %%CITATION = PTPKA,107,561;%%

\bibitem{Takagi}
  S.~Takagi,
  Prog.\ Theor.\ Phys.\  {\bf 109}  (2003), 233.
  %[arXiv:hep-ph/0210227].
  %%CITATION = PTPKA,109,233;%%


\bibitem{Ikeda:2001vc}
  T.~Ikeda,
  Prog.\ Theor.\ Phys.\  {\bf 107}  (2002), 403.
  %[arXiv:hep-ph/0107105].
  %%CITATION = PTPKA,107,403;%%

  Y.~Fueki, H.~Nakkagawa, H.~Yokota and K.~Yoshida,
  Prog.\ Theor.\ Phys.\  {\bf 110}  (2003), 777.
  %[arXiv:hep-ph/0212362].
  %%CITATION = PTPKA,110,777;%%

\bibitem{Nakkagawa:2007hu}
  H.~Nakkagawa, H.~Yokota and K.~Yoshida,
  %``Phase structure of thermal QED based on the hard thermal loop improved
  %ladder Dyson-Schwinger equation. A 'gauge invariant' solution,''
  in the Proceedings of the 2006 International Workshop 
  "The Origin of Mass and Strong Coupling Gauge Theories,'' Nagoya, Japan,
  2006, ed. M.~Harada, M.~Tanabashi and K.~Yamawaki 
  (World Scientific, Singapore, 2008), 
  p.~220
  [hep-ph/0703134.]
  %%CITATION = HEP-PH/0703134;%%

  H.~Nakkagawa, H.~Yokota and K.~Yoshida,
  %``Phase Structure of Thermal QCD/QED:A Gauge Invariant Solution of the HTL
  %Resummed Improved Ladder Dyson-Schwinger Equation,''
  in the Proceedings of the Nagoya Mini-workshop 
  "Strongly Coupled Quark-Gluon Plasma: SPS, RHIC and LHC,''
  Nagoya, Japan,
  2007, ed. C.~Nonaka and M.~Harada,
  p.~173
  [arXiv:0707.0929].
  %%CITATION = ARXIV:0707.0929;%%

  H.~Nakkagawa, H.~Yokota and K.~Yoshida,
  %``Analysis of the Phase Structure of Thermal QED/QCD through the HTL Improved
  %Ladder Dyson-Schwinger Equation --On the Gauge Dependence of the
  %Solution--,''
  arXiv:0709.0323 [hep-ph].
  %%CITATION = ARXIV:0709.0323;%%


\bibitem{Mitsutani:2007gf}
  M.~Kitazawa, T.~Kunihiro and Y.~Nemoto,
  %``Novel collective excitations and quasi-particle picture of quarks  coupled
  %with a massive boson at finite temperature,''
  Prog.\ Theor.\ Phys.\  {\bf 117}  (2007), 103.
  %[arXiv:hep-ph/0609164].
  %%CITATION = PTPKA,117,103;%%

%\bibitem{Kitazawa:2007ep}
M.~Kitazawa, T.~Kunihiro, K.~Mitsutani and Y.~Nemoto,
  %``Spectral properties of massless and massive quarks coupled with massive
  %boson at finite temperature,''
  Phys.\ Rev.\  D {\bf 77} (2008), 045034.
   %%CITATION = PHRVA,D77,045034;%%


\bibitem{Shuryak:2004cy}
  E.~V.~Shuryak and I.~Zahed,
  Phys.\ Rev.\ C {\bf  70}  (2004), 021901;
  %[arXiv:hep-ph/0307267].
  %%CITATION = PHRVA,C70,021901;%%

  Phys.\ Rev.\ D {\bf  70}  (2004), 054507.
  %[arXiv:hep-ph/0403127].
  %%CITATION = PHRVA,D70,054507;%%

\bibitem{Brown:2003km}
  G.~E.~Brown, C.~H.~Lee, M.~Rho and E.~Shuryak,
  %``The anti-q q bound states and instanton molecules at t >= T(C),''
  Nucl.\ Phys.\  A {\bf 740}  (2004), 171;
  %[arXiv:hep-ph/0312175].
  %%CITATION = NUPHA,A740,171;%%

  J.\ Phys.\ G {\bf 30}  (2004), S1275.
  %[arXiv:hep-ph/0402068].
  %%CITATION = JPHGB,G30,S1275;%%

  H.~J.~Park, C.~H.~Lee and G.~E.~Brown,
  %``The problem of mass: Mesonic bound states above T(c),''
  Nucl.\ Phys.\  A {\bf 763}  (2005), 197.
  %[arXiv:hep-ph/0503016].
  %%CITATION = NUPHA,A763,197;%%

  G.~E.~Brown, B.~A.~Gelman and M.~Rho,
  %``Matter Formed At The Bnl Relativistic Heavy Ion Collider,''
  Phys.\ Rev.\ Lett.\  {\bf 96}  (2006), 132301.
  %%CITATION = PRLTA,96,132301;%%

  G.~E.~Brown, C.~H.~Lee and M.~Rho,
  arXiv:nucl-th/0507011.
  %%CITATION = NUCL-TH/0507011;%%


%\cite{Weldon:1998ic}
\bibitem{Weldon:1998ic}
  H.~A.~Weldon,
  %``New mesons in the chirally symmetric plasma,''
  arXiv:hep-ph/9810238.
  %%CITATION = HEP-PH/9810238;%%

\bibitem{ac}
  F.~Marsiglio, M.~Schossmann and J.~P.~Carbotte, 
  Phys.\ Rev.\ B {\bf  37}  (1988), 4965.

%\cite{Baym:1992eu}
\bibitem{Baym:1992eu}
  G.~Baym, J.~P.~Blaizot and B.~Svetitsky,
  %``Emergence of new quasiparticles in quantum electrodynamics at finite
  %temperature,''
  Phys.\ Rev.\  D {\bf 46} (1992), 4043.
  %%CITATION = PHRVA,D46,4043;%%





\end{thebibliography}
\end{document}